# On the possibility of obtaining MOSFET-like performance and sub-60 mV/decade swing in 1D broken-gap tunnel transistors

Siyuranga O. Koswatta,[*,†] Steven J. Koester,[*,‡] and Wilfried Haensch[*]

* IBM T. J. Watson Research Center, 1101 Kitchawan Rd, Yorktown Heights, NY, 10598, USA

† Was also with, Midwest Institute for Nanoelectronics Discovery (MIND), University of Notre Dame, Notre Dame, IN, USA, at the time of this work.

Contact info – email: skoswatt@us.ibm.com, Tel.: 1-914-945-2926

‡ Now at, Department of Electrical and Computer Engineering, University of Minnesota, Minneapolis, MN, USA.

*Abstract* – Tunneling field-effect transistors (TFETs) have gained a great deal of recent interest due to their potential to reduce power dissipation in integrated circuits. One major challenge for TFETs so far has been achieving high drive currents, which is a prerequisite for high-performance operation. In this paper we explore the performance potential of a 1D TFET with a broken-gap heterojunction source injector using dissipative quantum transport simulations based on the nonequilibrium Green's function formalism, and the carbon nanotube bandstructure as the model 1D material system. We provide detailed insights into broken-gap TFET (BG-TFET) operation, and show that it can indeed produce less than 60mV/decade subthreshold swing at room temperature even in the presence of electron-phonon scattering. The 1D geometry is recognized to be uniquely favorable due to its superior electrostatic control, reduced carrier thermalization rate, and beneficial quantum confinement effects that reduce the off-state leakage below the thermionic limit. Because of higher source injection compared to staggered-gap and homojunction geometries, BG-TFET delivers superior performance that is comparable to MOSFET's. BG-TFET even exceeds the MOSFET performance at lower supply voltages ($V_{DD}$), showing promise for low-power/high-performance applications.

I. INTRODUCTION

Research on tunneling field-effect transistors (TFETs) has been resurgent in recent years because of the potential of TFETs to reduce power dissipation in integrated circuits [1-14]. The gate-controlled band-to-band-tunneling (BTBT) mechanism employed in TFET operation leads to subthreshold swings (*S*) below the 60mV/decade thermionic MOSFET limit at room temperature [15], allowing reduction in both the off-state leakage ($I_{OFF}$) and the power supply voltage ($V_{DD}$) while maintaining a comparable performance. Scaling $V_{DD}$ is particularly important since active power dissipation is proportional to $V_{DD}^2$ [15]. One major challenge for TFETs so far has been their limited on-current ($I_{ON}$) because of the presence of a tunneling barrier. Poor source injection in TFETs could also lead to "slow turn-on" in output characteristics [16], and enhance the influence of the drain on channel electrostatics that may lead to undesirable Miller capacitance effects [17, 18]. Furthermore, TFET subthreshold characteristics with *S* <

60mV/dec are obtained only at very small current levels, in a limited bias range [13]. All these facts pose great challenges for the use of TFETs in high-performance applications.

In order to improve the tunneling current, TFETs with *staggered-gap* heterojunction source injectors have been investigated, mainly in the $Si_xGe_{1-x}$-Si material system (in this paper, the staggered-gap barrier alignment is defined to have $0 \leq E_{G\text{-}eff} < E_G$ in Fig. 1(a)) [5, 10-12]. Even in this case, however, carrier transmission is impeded by the finite tunneling barrier, and the current drive is expected to be below the MOSFET limit. In order to further improve the drive current, a *broken-gap* heterojunction TFET (BG-TFET) geometry has been recently investigated using numerical device simulations ($E_{G\text{-}eff} < 0$ in Fig. 1(b)) [19-22]. The preliminary results in [19] showed that the BG-TFET can outperform both the staggered-gap TFET (SG-TFET) and the homojunction TFET (HJ-TFET) geometries, and deliver device currents comparable to MOSFET's, along with $S < 60$ mV/dec even in the presence of electron-phonon (e-ph) scattering. Here, we build upon [19] to provide detailed insights into BG-TFET operational physics including the influence of scattering which is essential to fully understand the subthreshold behavior of a broken-gap source injector, and present a comprehensive comparison of device performance to other TFET geometries and an equivalent 1D MOSFET.

It should be noted that [21, 22] has come to the conclusion that BG-TFETs will have $S \geq 60$ mV/dec similar to a thermionic MOSFET, which is different from our findings. We attribute this discrepancy to the non-optimal device structure with an underlapped source region used in [21, 22], along with the phenomenological treatment of scattering based on the Büttiker probe model whereby a series of Fermi reservoirs are connected along the length of the device to emulate local carrier equilibration. Here we, however, demonstrate the necessity of *nonequilibrium* transport simulations for the proper evaluation of BG-TFETs over traditional models based on equilibrium transport. Furthermore, having an extended source underlapped region as in [21, 22] provides additional carrier thermalization near the source injection region, causing the device to operate as a thermionic MOSFET in series with a heterojunction diode.

The remainder of the paper is organized as follows. Section II describes the simulation model based on the nonequilibrium Green's function (NEGF) formalism, and the microscopic treatment of scattering based on the self-consistent Born approximation employed here. Section III presents detailed operational physics of the 1D BG-TFET, and discusses the requirements for achieving $S < 60$mV/dec in these devices. Section IV then compares the 1D BG-TFET performance to devices with alternate heterojunction tunnel barrier alignments as well as an equivalent CNT based 1D MOSFET. Section V provides additional discussions on the $V_{DD}$ scalability of BG-TFETs for low-power applications and suggests possible material systems for the experimental realization of this device concept, followed by conclusions in Section VI.

## II. SIMULATION METHOD

In this study we use a carbon nanotube (CNT) as the model channel material because it facilitates the treatment of 1D limit for electronic bandstructure based on the atomistic tight-binding formalism [23, 24]. Also, CNTs possess small/direct bandgaps and small/symmetric effective masses in conduction and valence bands, that are similar to technologically relevant III-V materials [25] (see Section V.(b) for additional material suggestions). More importantly, advanced quantum transport simulation models based on the NEGF formalism [23] along with self-consistent device electrostatics and microscopic treatment of e-ph scattering have already been developed for CNT devices [26-28]. As mentioned earlier the detailed treatment of scattering is essential to properly evaluate the BG-TFET concept, and the exact material-specific bandstructure information is of secondary in importance. To our knowledge, because of the computational complexity, there has not been a simulation tool that can treat III-V based heterojunction devices of diameters ~ 10nm including rigorous bandstructure effects, and nonequilibrium transport with microscopic treatment of scattering, all of which are necessary to fully evaluate BG-TFETs. Hence, the CNT system provides a rigorous treatment of nonequilibrium transport in the presence of scattering at the 1D limit in order to critically evaluate the operational behavior of this novel TFET device concept.

A detailed account of the simulation procedure is described in [27], and here we summarize some important points. We use the $p_z$-tight-binding Hamiltonian ($H_{pz}$) for CNT electronic structure, and consider 1D carrier transport through the first conduction/valence subbands under mode-space approximation [24, 27]. Carrier scattering by all relevant phonon modes in CNTs [29] is considered under nominal conditions unless specified otherwise, but we take the e-ph scattering parameters (phonon energy, $\hbar\omega$, and deformation potential, $D_{e-ph}$) as free inputs to explore the influence of other scattering conditions as well. In this case, the device Green's function is given by, $G(E) = [EI - H_{pz} - \Sigma_S - \Sigma_D - \Sigma_{scat}]^{-1}$ where $I$ is the identity matrix, and the self-energy functions, $\Sigma_{S,D}$ and $\Sigma_{scat}$, are for coupling to the semi-infinite source/drain contacts, and to the phonon bath, respectively (the energy dependence, and the matrix form of the variables is implicit) [23, 27]. As shown in Fig 1, the heterobarrier at the source-channel junction is modeled by a shift in the midgap energy of $H_{pz}$ for one segment of the device (i.e. by shifting the midgap potential $U$ in eqn. (5) of [27] by an energy $(E_G - E_{G\text{-}eff})$ to the right side of the heterointerface). Now, in the self-consistent Born approximation for the treatment of e-ph scattering the device Green's function ($G$) and the scattering self-energy function ($\Sigma_{scat}$) are calculated self-consistently [27]. Here we consider one-phonon processes, and higher order interactions are captured only as sequential scattering events.

Figure 1(c) shows the modeled device geometry with high-k ($t_{ins}$ = 2nm, $k$ = 16) gate-all-around structure. The gate electrode is assumed to have a zero thickness with a fixed boundary potential on that surface. We use a (13,0) zigzag CNT with bandgap $E_G$ = 0.82eV, diameter $d_{CNT}$ = 1nm, and identical effective masses of $m_C^* = m_V^* \approx 0.08 m_0$ near the band edges. Gated channel length, $L_{ch}$ = 20nm, and doped source/drain regions of, $L_{S/D}$ = 15nm are used. For the *n*-type TFET simulated here, source (*p*-type) and drain (*n*-type) doping concentration of 0.4/nm can be compared with carbon atomic density of 122/nm, and a doping degeneracy of $\left(E_V^{source} - E_{FS}\right) \approx \left(E_{FD} - E_C^{drain}\right) \approx 60$meV is achieved. An *n*-type TFET operation is explored here, but *p*-type operation is equally possible. Aforementioned device parameters are used in all simulations unless specified otherwise.

### III. 1D Broken-Gap TFET Operation

Figure 2(a) shows the local density of states, $LDOS(x,E)$, plot for the 1D BG-TFET under typical off-state conditions ($LDOS$ is the energy-position resolved density of states throughout the device), where the broken-gap heterojunction near the source-channel interface is clearly seen. Under this biasing condition, the conduction band in the channel is pulled above the valence band in the source. Figure 2(b) is the energy-position resolved electron density spectrum, $G^n(x,E)$, which clearly shows the occupation of *LDOS* by the respective Fermi reservoirs of each contact. The gate modulation of the channel barrier position moves the broken-gap energy region near or below $E_{FS}$ in order to increase the source injection in the on-state. We discuss the operational characteristics of 1D BG-TFETs below.

*(a) Device Characteristics*

Figure 3 shows the transfer characteristics for the 1D BG-TFET ($E_{G\text{-}eff}$ = -100 meV) under ballistic and dissipative transport conditions (we examine e-ph scattering parameters corresponding to CNTs [29] and InAs [30]). First of all, large on-state currents of ~ 10µA/tube are obtained under all transport conditions. This can be easily understood by observing the energy-position resolved current density spectrum in Fig. 4(a), which clearly shows efficient source injection across the broken-gap heterojunction. On the other hand, in the subthreshold regime at the ballistic limit in Fig. 3, it is observed that $S$ < 60mV/dec can be obtained over many orders of magnitude change in $I_{DS}$. Under ballistic transport, source bandgap region can effectively filter out the high-energy thermionic injection even in a BG-TFET leading to steep turn-off. In our modeled device with $L_{ch}$ = 20nm, the value of $I_{OFF}$ (at $V_{GS}$ = 0.0V) under ballistic transport is mainly limited by direct source to drain tunneling through the channel barrier region.

More importantly, in Fig. 3, phonon scattering is observed to considerably modify the subthreshold operation depending on the scattering conditions. For the case of CNT scattering parameters a significant increase in the off-state leakage is observed. This is due to the phonon absorption assisted tunneling mechanism depicted in Fig. 4(b) [31]. In this case, the "shoulder" type feature seen in the subthreshold current under scattering is due to the large optical phonon energy in CNTs (~ 180meV) with strong e-ph coupling that allows inelastic tunneling paths well above the source valence band-edge energy, Fig. 4(b). Interestingly, for conventional semiconductor materials where phonon energies are much smaller (for example, InAs with $\hbar\omega_{OP} \approx$ 30meV), absorption assisted leakage paths are efficiently turned off in the subthreshold regime, and the "shoulder" type feature is suppressed in Fig. 3.

Nevertheless, even with CNT specific scattering parameters, the 1D BG-TFET in Fig. 3 still produces $S < $ 60mV/dec in a range of gate biases (0.0V $\leq V_{GS} \leq$ 0.1V), and a minimum spot-swing of $S \approx$ 21mV/dec. In this bias range with a $V_{GS}$ swing of 0.1V, $I_{DS}$ is reduced by ~ 800x which can be compared to only ~ 50x for $S = $ 60mV/dec thermionic MOSFET limit. The dashed-star curve in Fig. 3 produces $I_{OFF}$ ($V_{GS} = $ 0.0V) = 4.3x10$^{-4}$ µA/tube, $I_{ON}$ ($V_{GS} = $ 0.4V) = 8.1 µA/tube, and $I_{ON}/I_{OFF} = $ 1.9x10$^4$. This leads to, with parallel integration of ~ 230 tubes/µm, $I_{OFF} \approx$ 100 nA/µm and $I_{ON} \approx$ 1.9 mA/µm at $V_{DD} = $ 0.4V, all of which are very promising for low-power, high-performance circuit operation. Output characteristics ($I_{DS}$-$V_{DS}$) of BG-TFETs including the effect of additional series resistance have been discussed in [19], and it is not further addressed here.

*(b) Reasons for obtaining S < 60 mV/dec*

In this section we provide additional insights into the operational physics of BG-TFETs. It has been identified that there are two main reasons behind the observation of $S < $ 60mV/dec in 1D BG-TFETs under subthreshold conditions: 1) suppression of density of states (DOS) near the source-injection position, and 2) nonequilibrium carrier distribution below the thermionic limit in that region. Figure 5(a) is a zoomed-in view of the *LDOS* shown in Fig. 2(a) for the BG-TFET under off-state conditions. Here, a notch region near the source-channel junction appears where DOS is suppressed because of longitudinal confinement effects. This leads to a reduction in thermionic leakage current due to lack of available states near the source-injection position resulting in *S* < 60mV/dec behavior. Interestingly, in 2D or 3D device geometries, such suppression of DOS near the notch energy region is *not* possible because of the presence of transverse momentum states. This underscores a unique benefit of the 1D geometry for implementing the BG-TFET concept. We note, however, that unlike in CNTs, the large $m_V^*$ of the heavy-hole valence band of conventional III-V nanowires may increase the DOS in the *p*-type notch region in the source which is detrimental to desired device operation. Therefore, additional band-engineering methods (such as strain) could be used to lift up the light-hole band above the heavy-hole band in order to allow preferential occupation of the light-hole band, and thus regain the beneficial features of small $m_V^*$ (also see Section V.(b)). On the other hand, the small $m_C^*$ of III-V nanowires could still suppress the DOS on the channel side of the broken-gap junction in *n*-type TFETs (similar to Fig. 2(a)), or in the *n*-doped source side of *p*-type TFETs.

Figure 5(b) studies the influence of scattering on the electron distribution function (i.e. the occupation probability), $f_{dist}(x,E) = G^n(x,E)/LDOS(x,E)$, inside the notch region. Here, e-ph scattering with $\hbar\omega_{OP} \approx 30$meV optical phonons (similar to InAs) is considered, where the deformation potential is gradually increased from its nominal value ($D_{e-ph}$ = 3.7eV/Å [30]) in order to explore the sensitivity of the subthreshold operation to carrier scattering. In Fig. 5(b), in the absence of scattering (solid blue line), $f_{dist}$ is solely determined by the source and drain reservoir Fermi functions wherein the energy filtering effect of the source bandgap region is clearly seen. On the other hand, the presence of scattering inside the notch region leads to momentum and energy redistribution of carriers. As depicted in Fig. 5 (a), if the carrier scattering rate is slower than the propagation rate out into the channel ($1/\tau_{scat} \ll 1/\tau_{prop}$), a significant nonequilibrium distribution develops inside the notch region, leading to occupation probabilities below the thermionic limit, Fig. 5(b). Similar observations on the evolution of nonequilibrium distribution functions in the presence of scattering have been previously reported using Monte Carlo simulations [32], and "source starvation" effects have been predicted in the above-threshold operation of conventional MOSFETs [33]. Consequently, nonequilibrium effects can be present even in the subthreshold operation of BG-TFETs (Fig. 5(b)).

With increasing scattering rates (e.g. dotted red line in Fig. 5(b)), the distribution function inside the notch region rises above the ballistic limit (solid blue line), and eventually approaches the thermionic limit (indicated by the solid black line). In other words, under strong thermalization conditions the carrier distribution inside the notch region becomes "warmer" compared to the ballistic limit, and consequently, the off-state leakage current increases. Thus, in Fig. 5(c) the steep subthreshold slope gradually degrades with increasing scattering rates, and reaches the 60mV/dec thermionic limit under extremely high scattering conditions (note that the scattering rate, $1/\tau_{scat} \propto D_{e-ph}^2$, so the dotted red line corresponds to ~ 1600x increase in $1/\tau_{scat}$ from the nominal value). At large $D_{e-ph}$ values, $\tau_{scat}$ decreases, and thus the aforementioned condition ($1/\tau_{scat} \ll 1/\tau_{prop}$) does not remain valid at very high scattering rates. Also, Fig. 5(c) shows that the on current decreases at elevated scattering rates, as expected. We emphasize that the results under very high scattering rates are of a qualitative character because of the perturbative treatment of e-ph interactions.

It should be noted that electron-electron scattering is not treated in this study, and though this is one of the main thermalization processes in 2D and 3D, it is expected to be suppressed in 1D because of prohibitive energy-momentum conservation requirements [34]. Furthermore, other *elastic* scattering processes (such as Coulomb and surface-roughness scattering) are *not* expected to lead to carrier thermalization in 1D because of the pure backscattering nature of elastic scattering at the 1D limit. Thus, the subthreshold swing in 1D BG-TFETs is not expected to degrade by elastic scattering. Note that, even in 2D or 3D device geometries a sub-thermal distribution could develop near the broken-gap junction if the carrier scattering were lower, and lead to $S < 60$mV/dec operation. The gate-all-around device geometry should also reduce the electrostatic band bending distance, thereby decreasing the length of the notch region available for carrier thermalization. All of the above characteristics highlight the unique benefits of the 1D limit of device operation for BG-TFETs.

## IV. DEVICE COMPARISON

*(a) Transfer Characteristics*

Transfer characteristics ($I_{DS}$-$V_{GS}$) and their dependence on effective bandgaps ($E_{G\text{-}eff}$) for the modeled device structure are shown in Fig. 6(a) under dissipative transport with CNT specific phonon scattering. An equivalent CNT based 1D n-MOSFET is also included for comparison (with n-type source/drain doping of 0.9/nm to eliminate source exhaustion effects). It is observed that the above-threshold current significantly improves for the band alignment changing from homojunction to broken-gap heterojunction. Also, $S < 60$ mV/dec is obtained at higher overall current levels in the BG-TFET which is important to overcome any non-ideal leakage mechanisms (such as trap-assisted tunneling), and improves the device scalability due to higher immunity to direct source to drain tunneling leakage. In Fig. 6(a), even the staggered-gap geometry ($E_{G\text{-}eff} = E_G/2$) has significantly lower $I_{ON}$ compared to the broken-gap case ($E_{G\text{-}eff} = -100$meV). The $I_{ON}$ vs. $I_{OFF}$ plot in Fig. 6(b) clearly confirms a large increase in $I_{ON}$ (a right shift) with increasing heterobarrier discontinuity. On the other hand, the 1D MOSFET can deliver a larger $I_{ON}$ at the supply voltage considered here, $V_{DD} = 0.4$V (the effect of $V_{DD}$ scaling will be discussed in Section V.(a)). It is noted that, even though downstream transport beyond the tunnel junction is less important in conventional TFETs where the tunnel barrier dominates the overall device conductance [4, 5, 9], it can become a performance limiter in the high-injection broken-gap geometry. Therefore, a possible scaling path for BG-TFETs would include both the improvements in heterobarrier properties, as well as channel transport properties.

*(b) Device Performance*

Figure 7 plots the switching delay ($\tau_{switching}$) vs. $I_{OFF}$ for the devices in Fig. 6, calculated here by, $\tau_{switching} = (Q_{ON} - Q_{OFF})/I_{ON}$ where $Q_{ON}/Q_{OFF}$ are the total charge distributions throughout the device in the on/off states, respectively at $V_{DS} = V_{DD}$. Even though additional refinement for the delay estimate can be obtained by using an "effective drive current" model [35], the above relation is sufficient for the present discussion. Because of the fringe fields between the gate electrode and the source/drain regions, included in our 2D electrostatic solution, an "intrinsic" parasitic capacitance of, $C_{par} \approx 1$ aF/tube is present in all the devices. Figure 7 shows that the BG-TFET and the 1D MOSFET have similar performance even though the MOSFET delay can be improved at the expense of higher standby leakage. Furthermore, a significant increase in the switching delay with increasing heterobarrier overlap is seen, and the delay is substantially degraded for the HJ-TFET. This observation is in contrast to the previous predictions [16] that the gate delay of TFETs should *not* depend on the tunneling barrier properties.

We can gain further insights into the dependence of the gate delay by noting that $\tau_{switching}$ can be written as,

$$\tau_{switching} \approx \frac{(Q_{ch-S} + Q_{ch-D}) \cdot L_{ch} + Q_{par}}{I_{ON}} \qquad (1)$$

where $Q_{ch-S}$, $Q_{ch-D}$ are the source and drain injected charge inside the channel, respectively (Fig. 8) [36], and $Q_{par}$ is due to any parasitic or load capacitances that are inherent in circuits. Since $I_{ON} = Q_{ch-S} \cdot v_{ave}$, where $v_{ave}$ is the average carrier velocity inside the channel, (1) can be rewritten as,

$$\tau_{switching} \approx \left\{ 1 + \frac{Q_{ch-D}}{Q_{ch-S}} + \left( \frac{Q_{par}}{Q_{ch-S} \cdot L_{ch}} \right) \right\} \cdot \frac{L_{ch}}{v_{ave}} \qquad (2)$$

The first term in (2) corresponds to the intrinsic transit time of the device, while the second and third terms show that the influence of back-injected charge, and parasitic or load capacitances on the switching delay are minimized when $Q_{ch-S}$ is large. We emphasize that (2) is an important relation generally applicable to the performance analysis of any transistor, clearly showing key contributors to the overall delay. Figure 8 confirms the significantly high source injection in BG-TFETs compared to other heterobarrier alignments, as seen by the large source contribution ($C_{ch-S}$) to the channel capacitance ($C_{ch-tot}$). It is also seen that all TFETs can have a large drain-originating channel capacitance ($C_{ch-D}$) at higher gate biases [18, 36]. (The separation of the contributions from $C_{ch-S}$ and $C_{ch-D}$ to $C_{ch-tot}$ in TFETs has been described in [36]). From (2), the necessity of efficient source injection facilitated by broken-gap heterojunctions is clearly evident for high-performance applications.

## V. DISCUSSION

*(a) $V_{DD}$ Scalability*

Because of the $S < 60$mV/dec behavior in 1D BG-TFETs it is instructive to explore the possibility of lowering $V_{DD}$ to reduce active power dissipation in them, while still retaining high-performance. Figure 9(a) compares the above-threshold $I_{DS}$-$V_{GS}$ characteristics for the BG-TFET and the 1D MOSFET, under $V_{DD}$ scaling. Here, the subthreshold characteristics for the two devices are similar to that in Fig. 6(a), with an iso-$I_{OFF}$ (= $4.3 \times 10^{-4}$ µA/tube) condition at $V_{GS} = 0.0$V. In Fig. 9(a) it is clearly seen that the BG-TFET can deliver a higher drive current compared to the MOSFET under moderate gate overdrives for all $V_{DD}$ values. An interesting observation is that the TFET current can decrease (i.e. negative transconductance) at large $V_{GS}$ when $V_{DS}$ is small. This is due to the source depletion effect at large $V_{GS}$ that degrades the tunnel transmission at the source-channel junction in the current-carrying energy range (see, Fig. 4(a)), and this behavior is expected to be general to all TFETs.

The $I_{ON}$ vs. $I_{OFF}$ behavior in Fig. 9(b) shows that, when scaling $V_{DD}$, the BG-TFET can deliver a higher drive current compared to the MOSFET at smaller $I_{OFF}$ values. Furthermore, the reduction in $I_{ON}$ at a given $I_{OFF}$ value (say, $1 \times 10^{-3}$ µA/tube) under $V_{DD}$ scaling is much more gradual for the BG-TFET compared to the MOSFET. Therefore, it is expected that the BG-TFET can deliver a higher performance compared to the MOSFET, along with lower active power dissipation with $V_{DD}$ scaling. This is indeed confirmed in Fig. 10 where the switching delay ($\tau_{switching}$) and the switching energy ($E_{switching}$) for the two devices are compared. Here, the switching energy is calculated as, $E_{switching} = (Q_{ON} - Q_{OFF}) \cdot V_{DD}$, which is the energy required per on-off transition. In Fig. 10(a) it is clearly seen that the increase in $\tau_{switching}$ for the BG-TFET is very modest when $V_{DD}$ is scaled from 0.4V to 0.2V, while there is a substantial reduction in $E_{switching}$ in Fig. 10(b). On the other hand, the MOSFET shows a significant degradation in delay with $V_{DD}$ reduction.

Figure 11 further explores the $V_{DD}$ scalability of BG-TFETs compared to 1D MOSFETs, especially in the presence of additional parasitic/load capacitances. Here, in Fig. 11(a) the transistor performance ($1/\tau_{switching}$) is plotted, and in Fig. 11(b) $E_{switching}$ is plotted, both at $I_{OFF} = 10^{-3}$ µA/tube (i.e. a vertical line in Fig. 10 at $I_{OFF} = 10^{-3}$ µA). An additional charging time of $C_{par}V_{DD}/I_{ON}$, and an additional charging energy of $C_{par}V_{DD}^2$ have been considered in the presence of the extra capacitances. In Fig. 11(a) it is seen that the BG-TFET can deliver a higher performance even in the presence of realistic capacitive loads, and compared to the MOSFET, the performance does not degrade as strongly with $V_{DD}$ scaling. On the other hand, Fig. 11(b) clearly shows a significant reduction in switching energy with $V_{DD}$ scaling. Therefore, it is evident that the BG-TFET is a promising candidate to lower power dissipation in circuits while retaining high performance.

*(b) Material Choices*

Finally, it is noted that the 1D BG-TFET geometry could be implemented in nanowire based III-V heterostructures such as InAs-GaSb or InAs-InSb material systems [37, 38]. Large peak current and negative-differential resistance (NDR) behavior in broken-gap heterojunctions have already been demonstrated in bulk InAs-GaSb inter-band tunnel diodes [39, 40]. The latter behavior is especially promising because it shows the reduction in carrier conduction in the valley current region which indicates sub-thermal carrier distribution effects even in bulk broken-gap devices [39]. One main challenge for obtaining the broken-gap alignment in III-V nanowire geometries, however, is the influence of transverse energy confinement that tends to close the broken-gap and turn it into the staggered-gap configuration [20] (confinement also increases the band gap in the channel region which is beneficial for scaling and in reducing ambipolar conduction due to tunneling near the drain). In particular, the heavy-hole valence band requires very small transverse dimensions to attain large 1D confinement energies, but at these dimensions, the conduction band confinement on the other side of the heterojunction is too large to maintain the broken gap. One approach to mitigate this problem could be to use a smaller wire diameter in the source and a wider diameter in the channel [22, 38]. Strain engineering may also provide additional means to split the light-hole valence band above the heavy-hole. Nevertheless, detailed bandstructure calculations in III-V heterojunction materials (such as in [20]) will be required to rigorously explore such band engineering schemes to enable the 1D BG-TFET device concept.

## VI. CONCLUSION

This paper presented a detailed study of 1D BG-TFET geometry using dissipative quantum transport simulations based on the NEGF formalism. The broken-gap heterojunction significantly increases the source-injection efficiency in the on-state, delivering MOSFET-like device performance. 1D BG-TFETs can also allow significant reduction in circuit power dissipation, while retaining high performance. It is recognized that the 1D nature of transport leads to reduction in carrier thermalization and beneficial quantum confinement effects near the source-injection region, producing robust subthreshold operation with $S < 60$mV/dec in 1D BG-TFETs. Therefore, it is expected that the 1D broken-gap geometry will be able to overcome many performance challenges facing the traditional TFET designs, and should be experimentally demonstrable using III-V heterostructure nanowires.

## ACKNOWLEDGMENT

S.O. Koswatta thanks Prof. Patrick Fay, Prof. Alan Seabaugh, and Mr. Yeqing Lu of the University of Notre Dame for fruitful discussions, and Dr. Paul Solomon of IBM for a critical reading of the manuscript. Authors thank the NSF Network for Computational Nanotechnology (NCN) at Purdue University for computational support.

**List of Figure Captions:**

Fig 1. (a) Staggered-gap, and (b) broken-gap heterojunction alignment, and (c) the modeled 1D device with gate-all-around geometry and the heterojunction at the source-channel interface.

Fig. 2. (a) Local-density of states, and (b) energy-position resolved electron density distribution, of the 1D BG-TFET ($E_{G\text{-}eff}$ = -100meV) at $V_{GS}$ = -0.05V, $V_{DS}$ = 0.4V under dissipative transport ($\hbar\omega_{OP}$ = 30meV, $D_{e\text{-}ph}$ = 3.7eV/Å).

Fig 3. $I_{DS}$-$V_{GS}$ characteristics of the BG-TFET at $V_{DS}$ = 0.4V under ballistic and dissipative transport.

Fig. 4. Energy-position resolved current spectrum (log scale) of the 1D BG-TFET ($E_{G\text{-}eff}$ = -100meV) with dissipative transport (CNT phonon modes) at $V_{DS}$ = 0.4V and; (a) $V_{GS}$ = 0.4V (above-threshold), and (b) $V_{GS}$ = -0.05V (subthreshold). Phonon absorption-assisted tunneling dominates the TFET off-state leakage. Carrier thermalization in the drain by phonon emission is also observed.

Fig. 5. (a) Zoomed-in view of *LDOS* from Fig. 2(a) near the source-channel junction (in log scale for additional clarity), (b) The electron distribution function inside the notch region at $x$ = 14.8nm, and (c) $I_{DS}$-$V_{GS}$ characteristics, under different e-ph scattering

conditions. In (b), the equilibrium source Fermi function is also indicated (solid black line).

Fig. 6. (a) $I_{DS}$-$V_{GS}$ characteristics at $V_{DS}$ = 0.4V for 1D TFETs with different band alignment at the source-channel junction with CNT-like e-ph scattering. A CNT-based 1D MOSFET with a similar geometry is also included for comparison. (b) $I_{OFF}$ vs. $I_{ON}$ behavior determined with a constant gate bias window of $\Delta V_{GS}$ = $V_{DD}$ = 0.4V for the devices in (a).

Fig. 7. Switching delay [$\tau_{switching}$ = ($Q_{ON}$ - $Q_{OFF}$)/$I_{ON}$] vs. $I_{OFF}$, under dissipative transport (legend same as in Fig. 6). Here, $Q_{ON}$/$Q_{OFF}$ is the total charge throughout the device in the on/off states at $V_{DS}$ = $V_{DD}$ = 0.4V, and accounts for ~ 1aF/tube parasitic capacitance calculated from our 2D device geometry.

Fig. 8. Gate capacitance at the middle of the channel ($C_{ch-tot}$) vs. $V_{GS}$ under dissipative transport at $V_{DS}$ = 0.4V (legend same as in Fig. 6). $C_{ch-tot}$ is the gate capacitance per unit channel length at the middle of the channel. Source/drain charge injection into the channel ($Q_{ch-S}$/$Q_{ch-D}$), and the corresponding capacitive contributions ($C_{ch-S}$/$C_{ch-D}$) are depicted on right.

Fig. 9. Comparison of $V_{DD}$ scalability of BG-TFET vs. 1D MOSFET. (a) $I_{DS}$-$V_{GS}$ (linear) shows that the BG-TFET delivers higher drive currents under moderate gate overdrives even at lower $V_{DD}$. (b) $I_{OFF}$ vs. $I_{ON}$ comparison, determined with a constant gate bias

window of $\Delta V_{GS} = V_{DD}$, under different $V_{DD}$ operating conditions. Note that the subthreshold characteristics between the BG-TFET and the MOSFET are similar to that in Fig. 6(a) for all $V_{DD}$ conditions with iso-$I_{OFF}$ at $V_{GS} = 0.0$V.

Fig. 10. Comparison of $V_{DD}$ scalability of BG-TFET vs. 1D MOSFET (legend same as in Fig. 9). (a) Switching delay, and (b) Switching energy, with an intrinsic parasitic capacitance of $C_{par} \sim 1$aF/tube.

Fig. 11. Influence of $V_{DD}$ scaling on the, (a) transistor performance ($1/\tau_{switching}$), and (b) switching energy, at $I_{OFF} = 10^{-3}$ µA/tube, in the presence of additional parasitic capacitances. Switching energy reduction by scaling $V_{DD}$, while maintaining high performance, is possible in BG-TFETs even in the presence of realistic parasitic capacitances.

**List of Figures:**

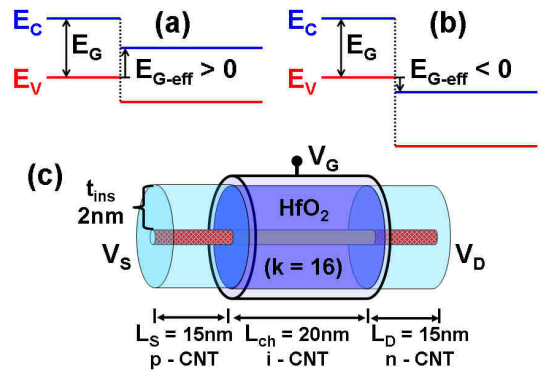

Fig. 1

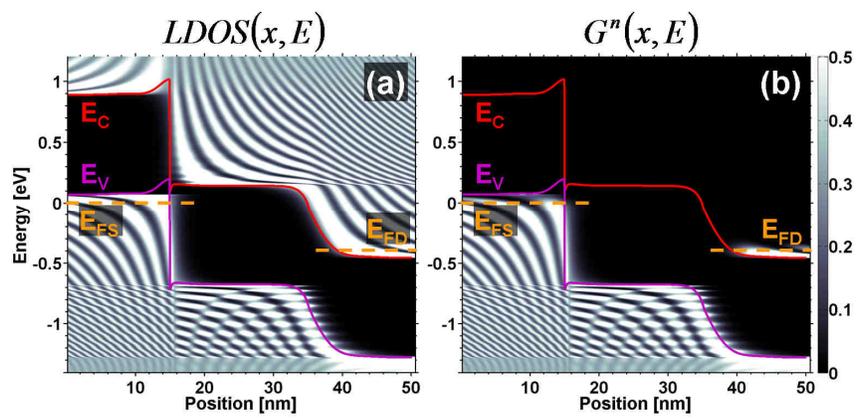

Fig. 2

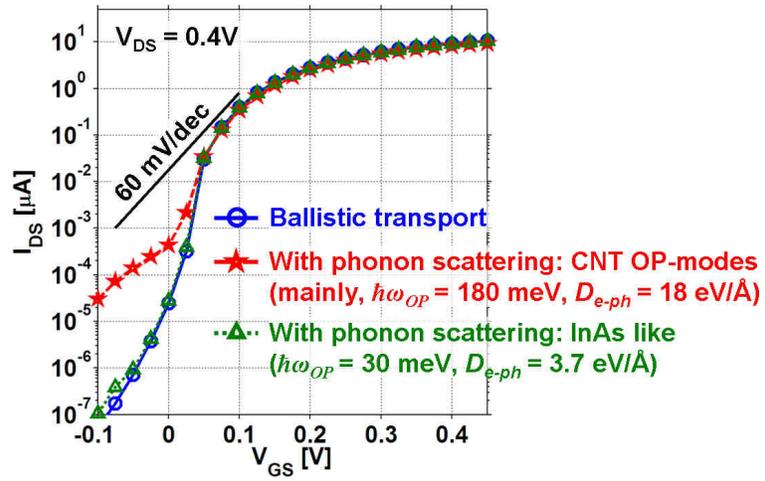

Fig. 3

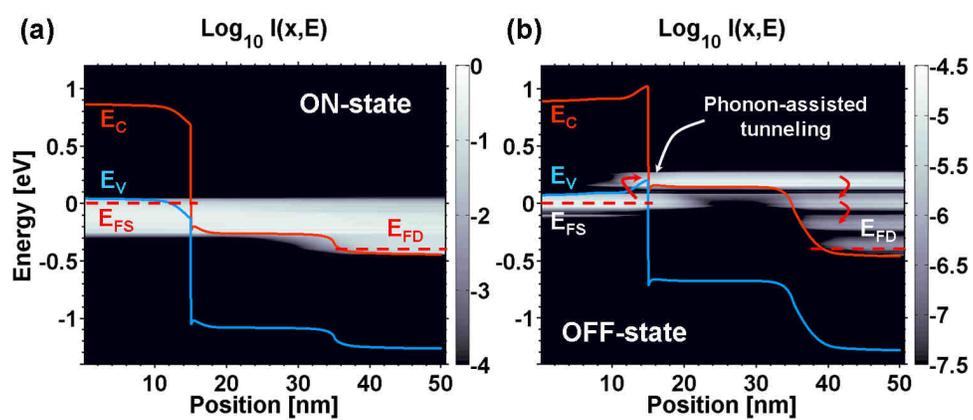

Fig. 4

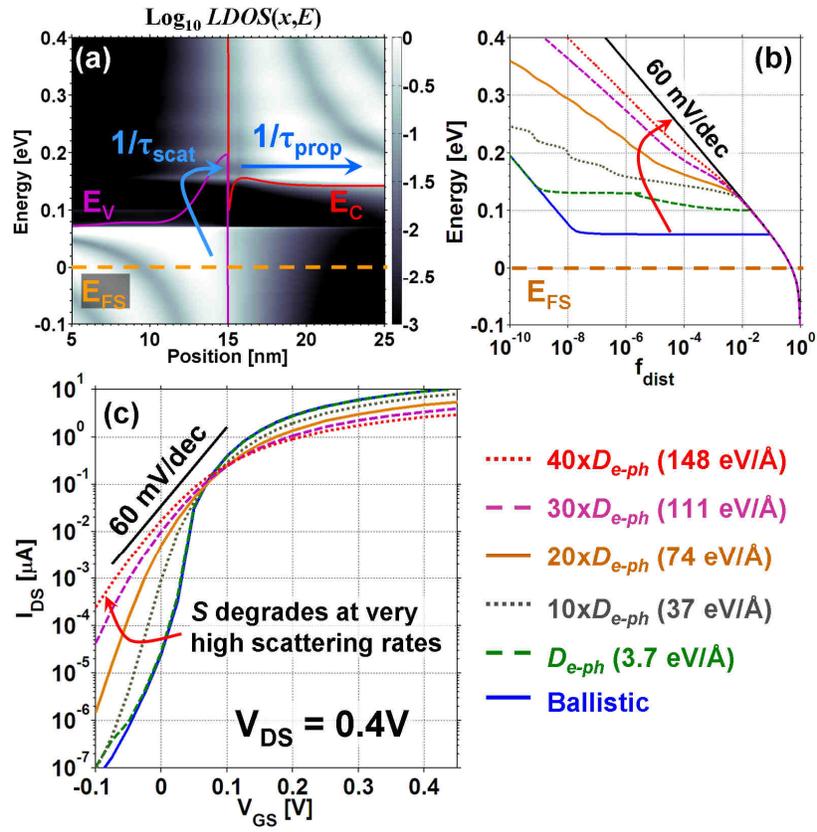

Fig. 5

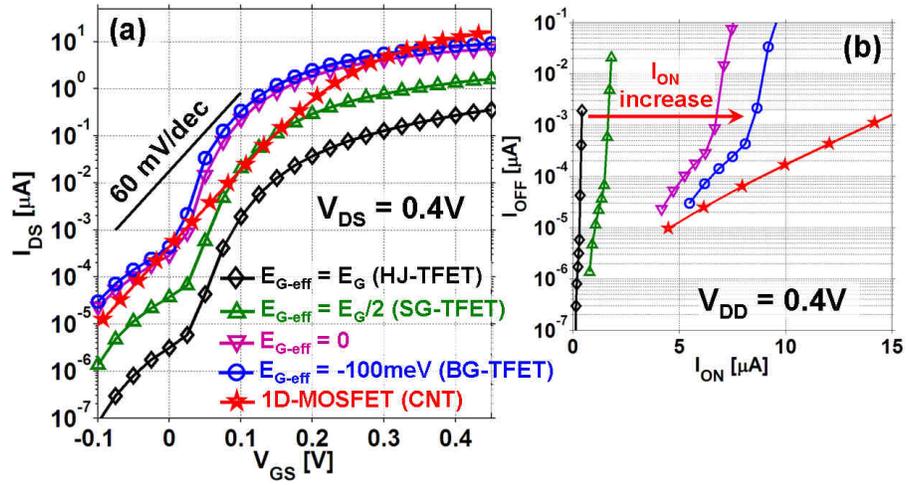

Fig. 6

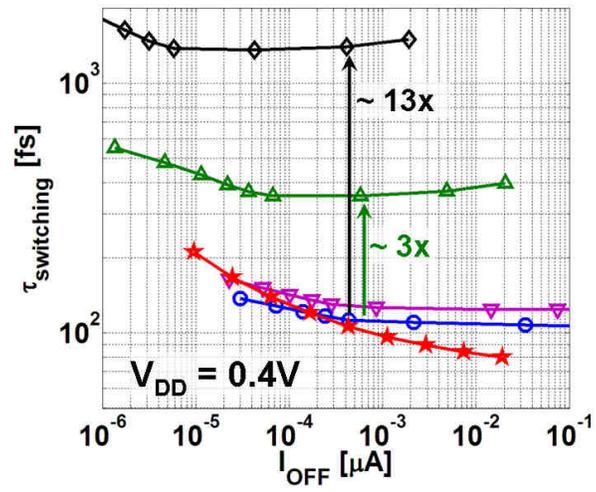

Fig. 7

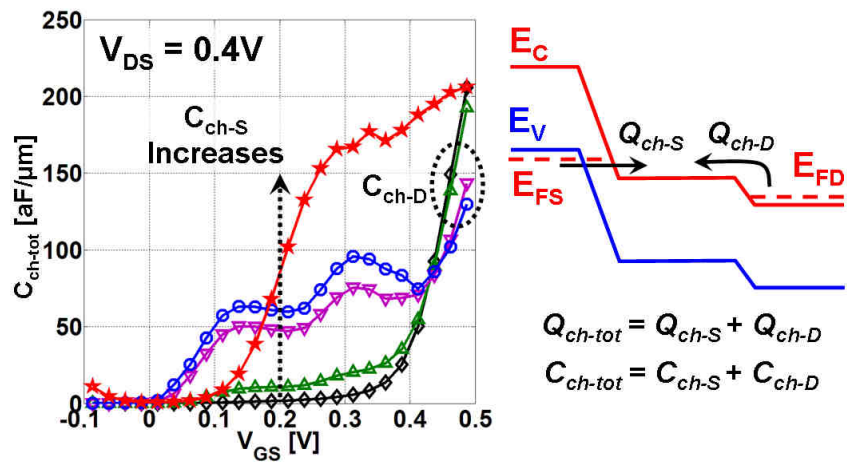

Fig. 8

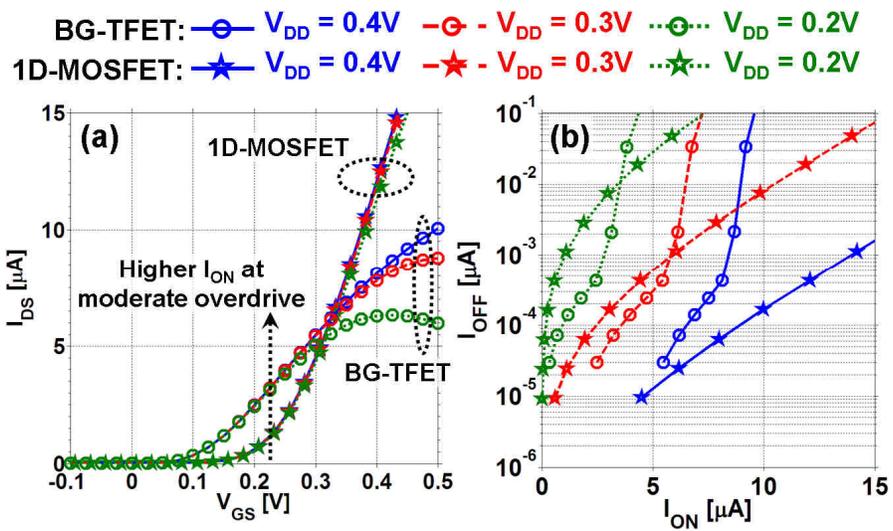

Fig. 9

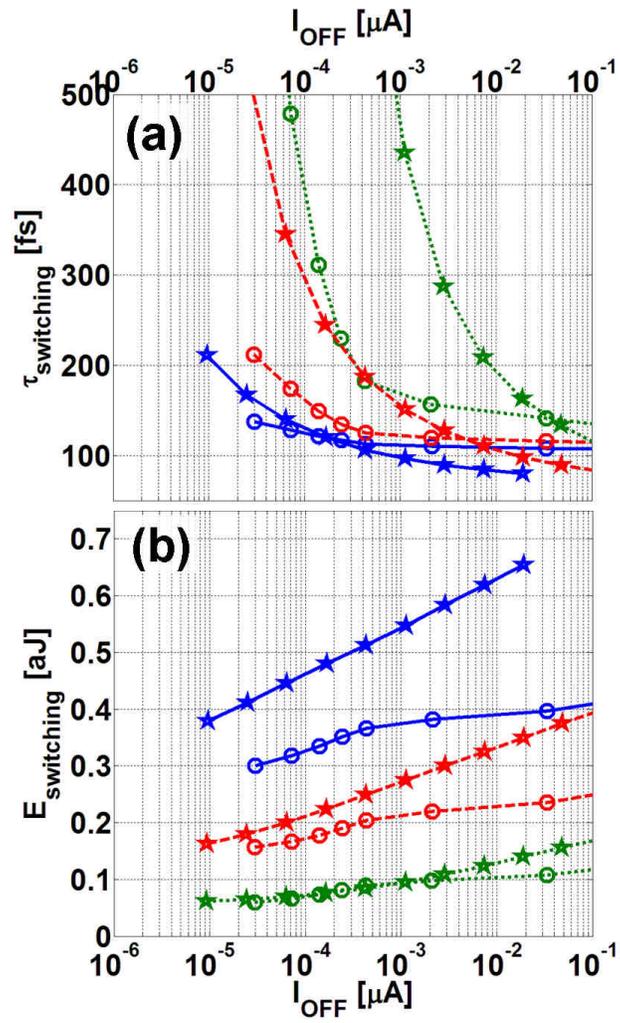

Fig. 10

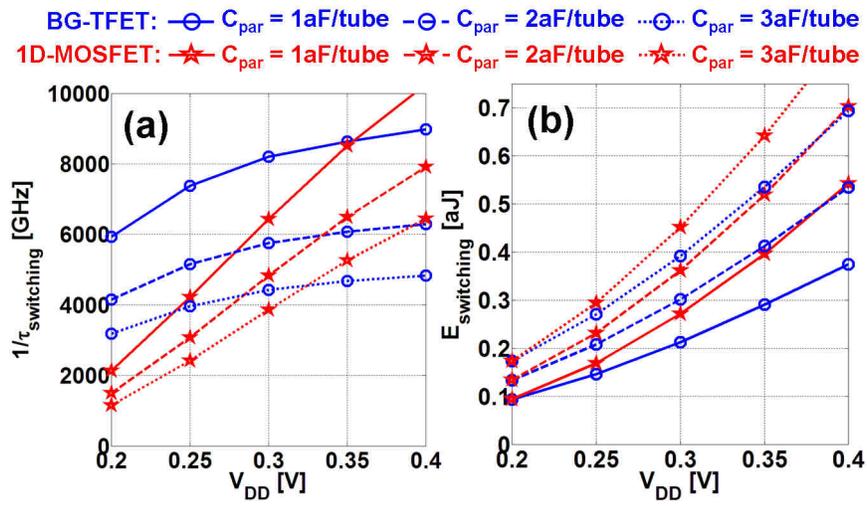

Fig. 11